\documentclass[english,superscriptaddress,showpacs,preprint]{revtex4-1}
\usepackage[T1]{fontenc}
\usepackage[latin9]{inputenc}
\usepackage{mathtools}
\usepackage{bm}
\usepackage{amsmath}
\usepackage{amssymb}
\usepackage{graphicx}


\usepackage{babel}
\begin{document}
\global\long\def\vc#1{\bm{#1}}%

\title{Positron scattering and transport in liquid helium}
\author{D. G. Cocks}
\affiliation{Research School of Physics, Australian National University, Canberra,
Australia}
\author{R. P. McEachran}
\affiliation{Research School of Physics, Australian National University, Canberra,
Australia}
\author{G. J. Boyle}
\affiliation{Linear Accelerator Technologies, Deutsches Elektronen-Synchrotron,
Hamburg, Germany}
\author{E. Cheng}
\affiliation{Research School of Physics, Australian National University, Canberra,
Australia}
\author{R. D. White}
\affiliation{College of Science, James Cook University, Townsville, Australia}
\begin{abstract}
In previous papers we have proposed a method for the \textit{ab initio}
calculation of fully differential cross-sections for electron scattering
in liquids and applied it to liquid argon, xenon and krypton. In this
paper, we extend the procedure to the consideration of positron scattering
in liquid helium, which is complicated by the annihilation process
as well as the fact that the electron definition for the region ``owned``
by a target atom used previously does not have a positron analogue.
We explore several physically motivated definitions to obtain effective
positron scattering in the dense fluid. We find that our calculations
of a pure helium system cannot precisely match experimental measurements,
however by including a small admixture (<0.1\%) of an impurity, we
can obtain reasonable agreement in the dense gas phase. In contrast,
our calculations do not match well to the liquid phase measurements.
This provides motivation to explore further multiple scattering effects
in the theory.
\end{abstract}
\maketitle

\section{Introduction}

Positrons are used in a variety of diagnostic applications, including
the medical diagnostic of PET (positron emission tomography), materials
analysis through PALS (positron annihilation lifetime spectroscopy)
and DBS (Doppler broadening spectroscopy) \citep{Charlton2001}. Positrons
can even be used as an indirect probe of the structure of the Galaxy~\citep{Prantzos2011}.
These experimental techniques generally rely on the interpretation
of gamma rays emitted from the annihilation of the positrons with
electrons.

To be able to interpret these diagnostics, it is essential to understand
how the positron propagates through the material under investigation.
As the concentration of positrons is typically very low, this falls
under the umbrella of swarm modelling \citep{RobsonBook}. In gases
this is usually explored through kinetic theory simulations, which
allows for a simple scaling with density for transport properties
such as annihilation rates and drift velocities. As the density significantly
increases, these scaling behaviours have historically been used directly,
even for systems as dense as liquids where the scaling laws break
down.

To model charged particle transport in liquids and dense gases, we
must account for effects such as multiple scattering and interaction
screening, using correlations between particles in the fluid. This
was first described by Lekner~\citep{Lekner1967a} and we have since
extended the procedure to calculate more accurate, ab-initio, fully-differential
effective elastic cross sections for electrons propagating in liquid
argon, xenon and krypton \citep{Boyle2015,Boyle2016,White2018} using
only the pair correlator for each fluid.

In this article, we investigate positron transport. On the one hand,
this should share much of the same properties of electron transport
through a fluid, as both the electron and positron are a light charged
particle. On the other hand, the interaction of the positron with
a single atom of the fluid is very different: it has no exchange interaction,
the sign of the Coulomb interaction is reversed and loss processes
are always present, even as the collisional energy approaches zero.
Note, however, that the polarisation interaction is similar for both
the positron and electron, as the induced dipole-charge interaction
is independent of the sign of the charge.

An important feature of our approach is that the effective cross sections
are calculated in an \emph{ab initio} manner from an interaction potential.
This is useful because, a) less is known about the positron elastic
and annihilation cross sections as measurements are more difficult
than the corresponding electron system, and b) it is the interaction
between the charged particle and the atom that is modified in the
fluid, whereas the isolated-atom cross sections are not so simply
related to the effective cross sections in the fluid.

The structure of this article is as follows. We first describe the
methods that allow us to obtain cross sections in the gas and dense
fluid phases from scattering calculations and then how we can use
these to obtain the transport coefficients in the gas and dense fluid
phases. Comparison of calculated transport coefficient calculations
under equilbirum and non-equilbrium conditions (driven out of equilibrium
through the application of an applied field) with available experimental
measurements represents a stringent test on the accuracy of our position
cross-sections for dense gas and liquid phases. We validate our scattering
calculations using gas phase data, for which both total cross section
measurements and transport data are available \citep{Mizogawa1985,Sullivan2008}.
Then we apply the dense fluid formalisms, for which we can compare
to experimental measurements in the dense gas \citep{Davies1989}
and liquid \citep{Pepe1995} regimes. The dense gas comparisons suggest,
with reference to previous analysis \citep{Boyle2014}, that there
is an incompatibility with several of the measurements. We have been
able to show that density effects are significant in the dense gas
phase only at low reduced electric fields, by performing full calculations
and through simple qualitative arguments. This has allowed us to suggest
that an admixture of an impurity may resolve the discrepancies between
our calculations and experimental measurements. We will then perform
similar analysis for the liquid phase, and discuss the incompatibilities
between the calculations and measurements.

\section{Kinetic theory and transport properties}

The kinetic equation used here to describe a positron swarm subject
to an external electric field $\boldsymbol{E}$ in a background of
gaseous or liquid helium is Boltzmann's equation (BE) for the phase-space
distribution function. As shown in our previous works \citep{Boyle2016,Boyle2015},
comparison with positron swarm experiments can be made with only the
steady-state, spatially-homogeneous solution: 
\begin{equation}
\frac{q\boldsymbol{E}}{m}\cdot\frac{\partial f}{\partial\boldsymbol{v}}=-J(f),\label{eq:BoltzSS}
\end{equation}
by performing a Legendre polynomial $P_{l}$ decomposition of the
distribution function:
\begin{equation}
f(\boldsymbol{v})=\sum_{l=0}^{\infty}\,f_{l}(\epsilon)P_{l}(\mu),\label{eq:LegExpan}
\end{equation}
and the collision integral $J(f)$:
\begin{equation}
J(f)=\sum_{l=0}^{\infty}\,J^{l}(f_{l}).\label{eq:LegExpan-collint}
\end{equation}
Details of our calculation method can be found in \citep{Boyle2016,Boyle2015},
which include a specialized collision operator for the coherent elastic
scattering. For the current investigation, we must also include the
annihilation process for the positron. This requires the definition
of the annihilation collision operator 
\begin{equation}
J_{l}^{^{an}}(f_{l})=\nu_{an}(\epsilon)f_{l}
\end{equation}

The BE allows a connection between microscopic scattering information,
and macroscopic transport properties. The macroscopic transport quantity
of interest in this work is the average annihilation rate $\alpha_{an}$,
which can be calculated from the energy distribution function, $f_{0}(\epsilon)$,
via \citep{RobsonBook}

\begin{equation}
\alpha_{an}=2\pi\left(\frac{2}{m}\right)^{\frac{3}{2}}\int_{0}^{\infty}\epsilon^{\frac{1}{2}}\nu_{an}\left(\epsilon\right)f_{0}\left(\epsilon\right)d\epsilon\label{eq:annihilationRate}
\end{equation}

\section{Scattering of positrons by individual helium atoms}

The theoretical procedures used in this paper to describe the elastic
scattering of positrons from helium atoms, at energies below the positronium
formation threshold at $17.79~$eV, are given in \citep{McEachran2019}
and are essentially the same as those used in \citep{Boyle2016,Boyle2015}
for electron scattering from argon and xenon. Thus, only a brief discussion
of the overall method will be given here.

In the purely elastic energy region, only the static and polarization
potentials need to be included in the interaction for positron scattering.
The scattering of the incident positrons, with wavenumber $k$, by
helium atoms can then be described in the gaseous phase by the integral
equation formulation of the partial wave Dirac-Fock scattering equations
(see \citep{McEachran2019} for details). In matrix form, these equations
can be written as

\begin{equation}
\begin{pmatrix}f_{\kappa}(r)\\
g_{\kappa}(r)
\end{pmatrix}=\begin{pmatrix}v_{1}(kr)\\
v_{2}(kr)
\end{pmatrix}+\frac{1}{k}\int_{0}^{r}\mathrm{d}x\,G(r,x)\,\biggl[U(x)\begin{pmatrix}f_{\kappa}(x)\\
g_{\kappa}(x)
\end{pmatrix}\biggr]\label{eq:DiracFock}
\end{equation}
where $f_{\kappa}(r)$ and $g_{\kappa}(r)$ are the large and small
components of the scattering wavefunction, $G(r,x)$ is the free particle
Green's function and $U(r)$ is the local potential. In particular,
$U(r)$ contains the static as well as the dipole and quadrupole polarization
interactions, with the latter being calculated by the the polarized
orbital method \citep{McEachran1977,McEachran1977corr}. The calculation
of the momentum transfer cross section $\sigma_{\mathrm{mt}}$ from
these potentials is discussed in \citep{Boyle2015,Boyle2016}. 

For positron scattering we also require a cross section for annihilation
or its equivalent designation in terms of $Z_{\mathrm{eff}}$, the
effective number of atomic electrons \citep{Charlton1999}:
\begin{equation}
\sigma_{A}=\frac{\pi r_{0}^{2}c}{v}Z_{\mathrm{eff}}
\end{equation}
where $r_{0}$ is the classical electron radius, $c$ is the speed
of light ($c=1/\alpha$ in a.u. where $\alpha$ is the fine-structure
constant), $v$ is the velocity of the incident positron and
\begin{equation}
Z_{\mathrm{eff}}=\sum_{i=1}^{N}\int\left|\Psi(\bm{r}_{1},\bm{r}_{2},\ldots,\bm{r}_{N};\bm{x})\right|\delta(\bm{r}_{i}-\bm{x})\mathrm{d}\bm{r}_{1}\mathrm{d}\bm{r}_{2}\ldots\mathrm{d}\bm{r}_{N}
\end{equation}
Here $\Psi$ is the total scattering wavefunction and the $\bm{r}_{i}$
are the coordinates (including spin) of the atomic electrons while
$\bm{x}$ is the position vector of the incident positron. The quantity
$Z_{\mathrm{eff}}$ can then be expressed as
\begin{equation}
Z_{\mathrm{eff}}=Z_{\mathrm{eff}}^{0}+Z_{\mathrm{eff}}^{1}
\end{equation}
where
\begin{equation}
Z_{\mathrm{eff}}^{i}=\frac{1}{2\pi}\sum_{\kappa}\int_{0}^{r_{m}}\mathrm{d}r\left[\frac{f_{\kappa}^{2}(r)+g_{\kappa}^{2}(r)}{r^{2}}\right]\rho_{i}(r).\label{eq:Zeff_01}
\end{equation}
Here $\rho_{0}(r)$ is the unperturbed density of the atomic electrons
and $\rho_{1}(r)$ is the first-order correction. In terms of the
atomic wavefunctions $\rho_{0}(r)$ is given by
\begin{equation}
\rho_{0}(r)=\sum_{n\kappa}q_{n\kappa}\left[P_{n\kappa}^{2}(r)+Q_{n\kappa}^{2}(r)\right]
\end{equation}
where $P_{n\kappa}(r)$ and $Q_{n\kappa}(r)$ are the large and small
radial components of the atomic wavefunctions while $q_{n\kappa}=2|\kappa|$
is the occupation number of the $n\kappa$ subshell of a closed shell
atom.

The first-order charge density was determined by the non-relativistic
polarized orbital method \citep{McEachran1977,McEachran1978}, as
relativistic effects are essentially negligible in light atomic systems.
In the polarized-orbital method the first-order radial distortion
$F_{nl}^{\nu\nu^{\prime}}(r,x)$ of each atomic orbital $P_{nl}(r)$
is calculated adiabatically in the field of a point charge at a series
of fixed points $x$ (c.f. equation (12) of \citep{McEachran1977corr}).
The corresponding non-relativistic scattering wavefunction $f_{l}(r)$
is normalized at infinity according to
\begin{equation}
f_{l}(r)\sim\frac{\left[4\pi(2l+1)\right]^{\frac{1}{2}}}{k}\sin\left[kx-\frac{l\pi}{2}+\delta_{l}\right].
\end{equation}
Here, $k$ is the wavenumber of the incident positron while $\delta_{l}$
is the partial wave phase shift. The correction to the charge density
is then found by keeping only terms to first order and is given by
\begin{equation}
\rho_{1}(r)=\sum_{nl}q_{nl}\sum_{\nu\nu^{\prime}}(2\nu^{\prime}+1)\left(\begin{matrix}\nu & \nu^{\prime} & l\\
0 & 0 & 0
\end{matrix}\right)^{2}P_{nl}(r)F_{nl}^{\nu\nu^{\prime}}(r,r)
\end{equation}
where $q_{nl}=2(2l+1)$ is the occupation number of the $nl$ subshell
of a closed shell atom.

\begin{figure}
\begin{centering}
\includegraphics[width=0.9\columnwidth]{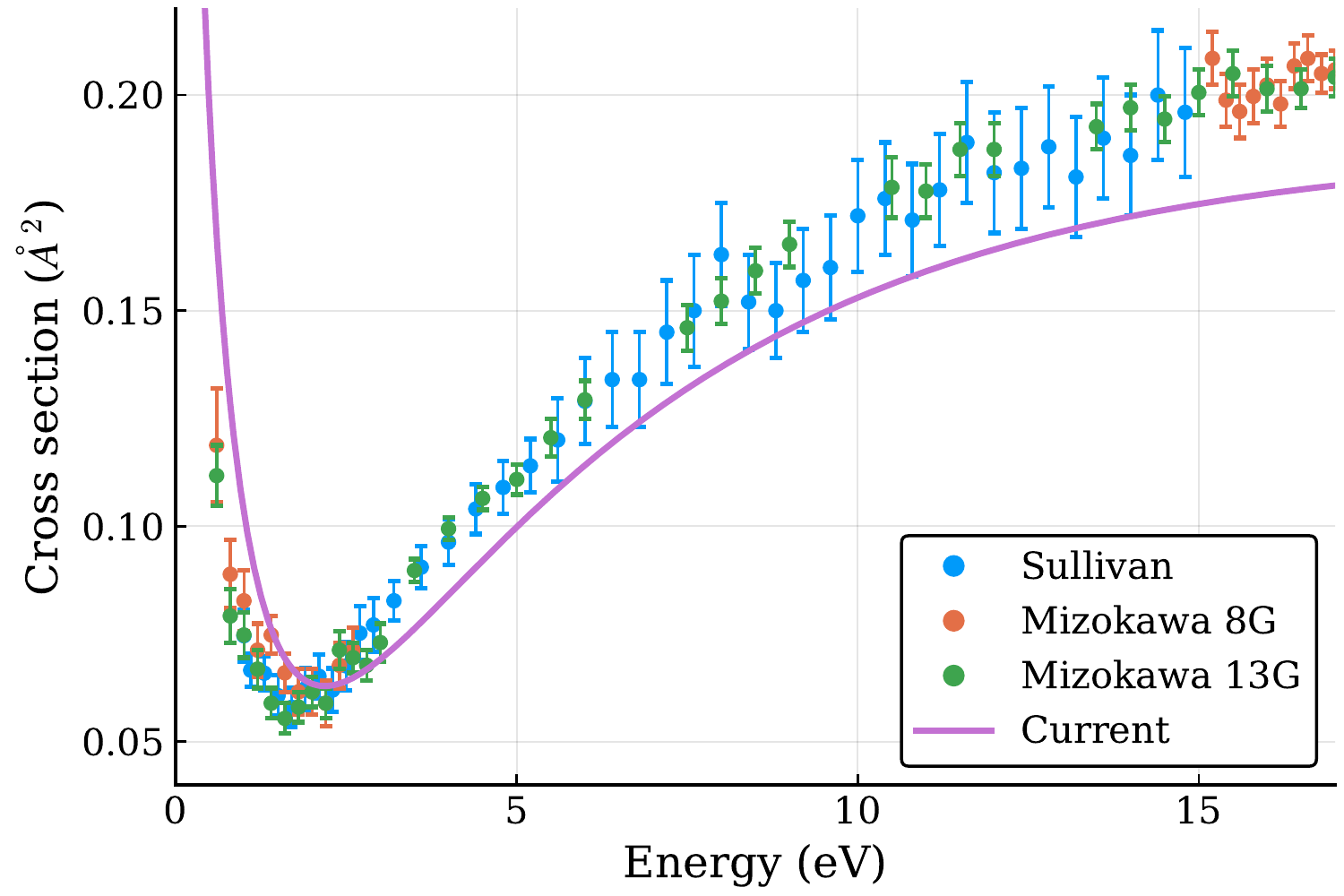}
\par\end{centering}
\caption{\label{fig:gas_cs}Comparison of our calculated gas phase cross sections
\citep{McEachran2019} used in this paper to various experimental
measurements \citep{Mizogawa1985,Sullivan2008}.}
\end{figure}

A comparison of the single-atom elastic cross sections to single-scatter
experiments is shown in figure~\ref{fig:gas_cs}.

\subsection{Transport coefficients}

\begin{figure}
\begin{centering}
\includegraphics[width=0.9\columnwidth]{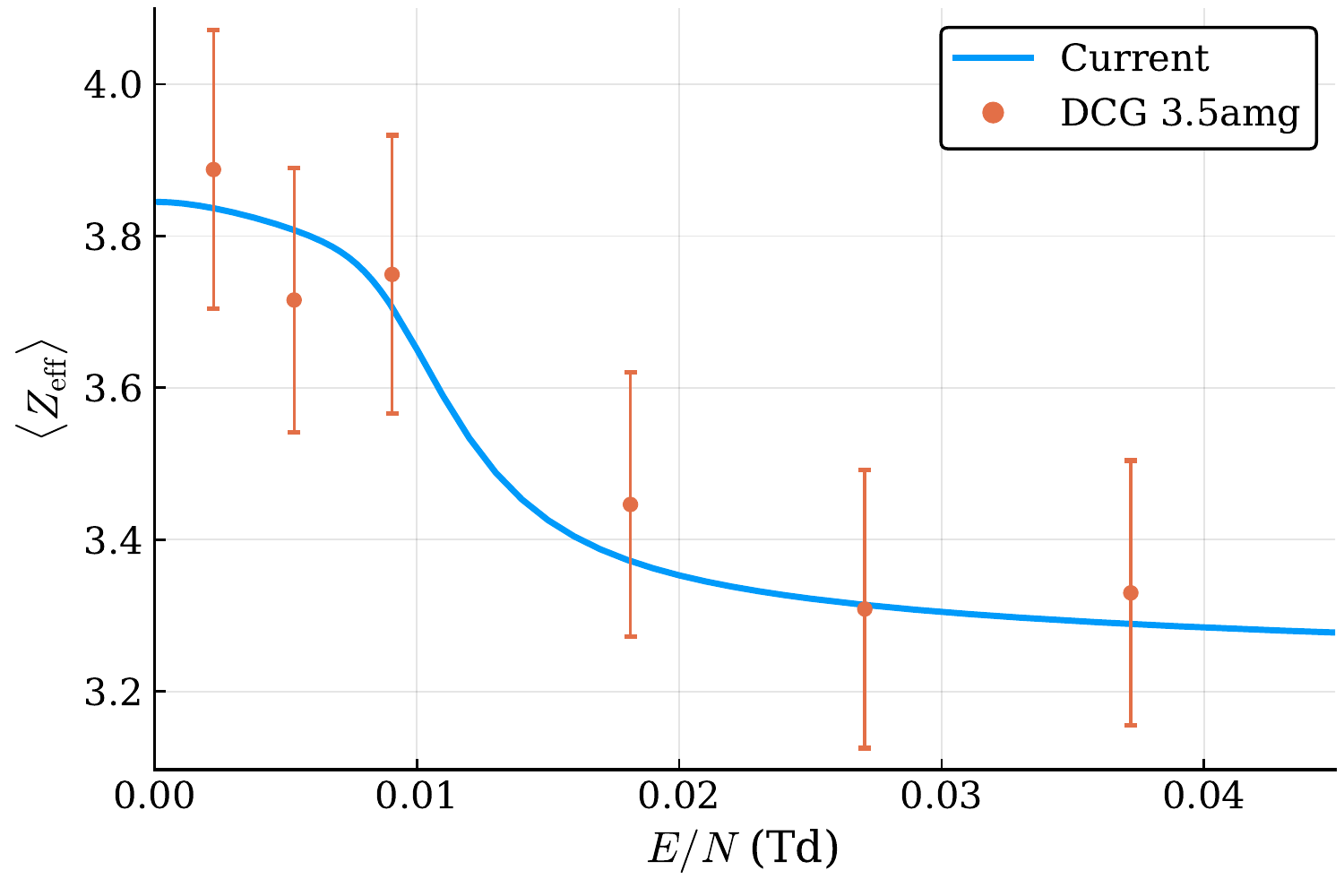}
\par\end{centering}
\caption{\label{fig:gas-Zeff-comp}The averaged $\langle Z_{\mathrm{eff}}\rangle(E)$
for gas phase compared to measurements for two different gas densities
\citep{Davies1989}. Although the calculation is within the error
bars of the measurements at 3.5~amg density, the higher density measurements
are significantly different.}
\end{figure}

In order to test our calculation procedure in the dilute gas case,
we can compare to various experimental measurements of the thermal
zero-field annihilation rate \citep{Paul1957,Charlton2001,Charlton2009}
and to field-dependent measurements at 3.5~amagat of Davies~\emph{et
al. }\citep{Davies1989}. The general consensus of the zero-field
effective atomic number for room temperature is $\langle Z_{\mathrm{eff}}\rangle_{T_{0}}\approx3.9$
and our value of $\langle Z_{\mathrm{eff}}\rangle_{T_{0}}=3.84$ at
300~K is in good agreement. Our field-dependent results, shown in
figure~\ref{fig:gas-Zeff-comp}, are also in agreement with experiment,
although the large uncertainties provide some leeway for variation.

We should note that the steady-state distribution, $f(\bm{v})$ in
equation~(\ref{eq:LegExpan}), is a non-equilibrium distribution,
even in the non-equilibrium case owing to the ``hole-burning'' effect
provided by the energy-dependence of the annihilation collision frequency.
Furthermore, it is also conceivable that the time-dependent behaviour
of the positron swarm, as it approaches steady-state, could result
in too few positrons that survive to reach the true steady-state $f(\bm{v})$
distribution. If this were the case, then the experimental measurements
would correspond to an average over transient distributions instead
of steady-state. Fortunately, it has been shown \citep{Green2017,Boyle2014}
that enough positrons survive to accurately represent the steady-state
distribution.

\section{Scattering of positrons by dense helium fluids}

Our approach to calculating the transport through liquids and dense
gases, referred henceforth as dense fluids, is presented in \citep{Boyle2015,Boyle2016}.
In these papers we detailed the procedure, originally proposed by
\citep{Cohen1967} for constructing effective scattering potentials
for electrons in dense media. The procedure is almost identical for
positron scattering and we do not repeat the formalism here but describe
only the changes we have made for the current application to positrons.
These include a) a contribution to the annihilation cross section
from the average over surrounding atomic charge densities, b) a different
choice of the outer radius of the scattering calculation and c) a
potential shift, similar to that applied in our investigation of liquid
krypton \citep{White2018}.

\subsection{The averaged electron density $\rho_{\mathrm{eff}}$}

Analogous to the effective total potential, one can define an effective
charge density with contributions from both the target atom and an
ensemble average contribution from the atoms in the bulk, which acts
to increase the positron annihilation rate in dense systems:

\begin{align*}
\rho_{\mathrm{eff}}(R) & =\rho_{L}(R)+\rho_{S}(R)\\
 & =\rho_{L}(R)+\frac{2\pi n}{R}\int_{r_{m}}^{\infty}ds\ sg(s)\int_{\left|R-s\right|}^{R+s}dt\,t\rho_{L}\left(t\right).
\end{align*}
Here $\rho_{L}=\rho_{0}+\rho_{1}$ corresponds to the focus atom's
charge density and $\rho_{S}$ denotes the surrounding average. Note
that the $r_{m}$ lower limit on the outer integral of $\rho_{S}$
indicates that only the charge density outside the region owned by
the target atom contributes to the averaged density of its surrounding
atoms. This is complementary to the upper limit of $r_{m}$ in equation~(\ref{eq:Zeff_01}).
In other words, we consider any charge density within a range $r_{m}$
of an atom to be ``owned'' by that atom; this is necessary to prevent
``double counting'' of the electrons for each atom. In the dilute
gas limit $r_{m}\rightarrow\infty$ and $\rho_{S}\rightarrow0$ as
required.

With the total averaged charge distribution defined, we can easily
extend the definition of $Z_{\mathrm{eff}}$ to include the total
contribution from the focus and surrounding atoms:
\begin{align*}
Z_{\mathrm{eff}} & =\frac{1}{n}\int dR\ \left(\rho_{L}(R)+\rho_{S}(R)\right)\left|\Psi(R)\right|^{2}\\
 & =Z_{\mathrm{eff}}^{L}+Z_{\mathrm{eff}}^{S}.
\end{align*}
We have found that in our current focus of helium, the contribution
of $Z_{\mathrm{eff}}^{S}$ to the total $Z_{\mathrm{eff}}$ is negligible,
however this may not be true for larger atoms. 

\subsection{Choice of $r_{m}=r_{\mathrm{WS}}$}

\begin{figure}
\begin{centering}
\includegraphics[width=0.9\columnwidth]{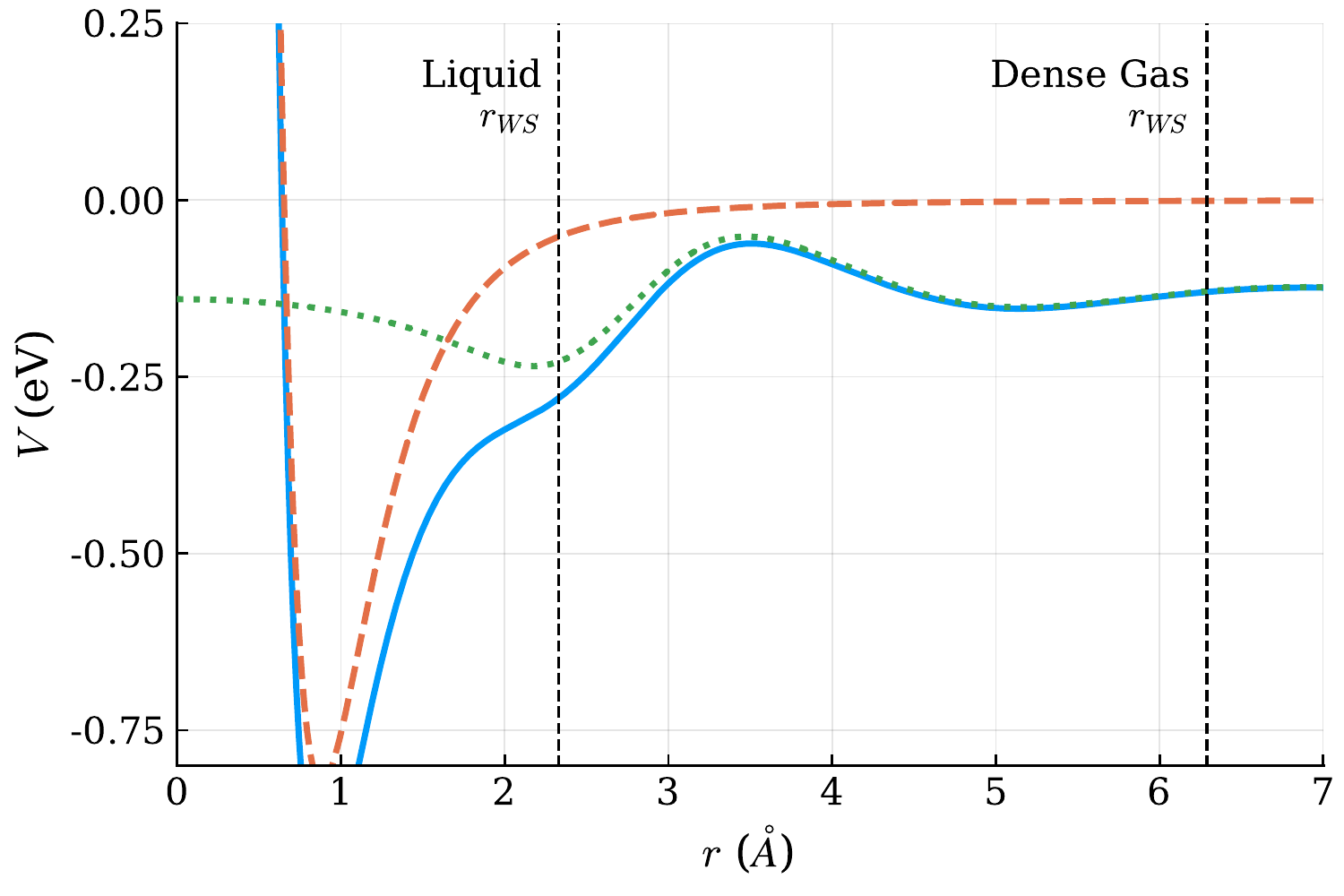}
\par\end{centering}
\caption{\label{fig:potentials}The potential due to the focus atom (dashed
orange), surrounding atoms (dotted green) and total scattering potential
(solid blue) in the case of scattering in the liquid. The value of
$r_{\mathrm{WS}}$ is shown as a black dashed vertical line. For comparison
the dense gas phase $r_{\mathrm{WS}}$ is also shown, however the
surrounding and total potential are different in the dense gas phase.}
\end{figure}

The value of $r_{m}$ represents the region of space ``owned'' by
the focus atom and distinguishes it from the rest of the bulk. In
comparison to our previous works involving electrons in dense fluids,
a different definition for $r_{m}$ is required for positrons in dense
fluids. In our previous works, we followed the lead of Lekner~\citep{Lekner1967a}
to choose $r_{m}$ as a turning point of the potential. However, in
the current case of positron scattering, where the sign of the static
potential is reversed, this definition results in a much larger value
of $r_{m}$ which appears to be physically invalid. Hence we make
an alternative choice of setting $r_{m}$ to the Wigner-Seitz radius,
$r_{\mathrm{WS}}=(4\pi N/3)^{-1/3}$. In the dense gas phase of helium
at $N=35.7$~amagat, this is $r_{\mathrm{WS}}=6.29\,\text{Å}$ and
in the liquid phase at $N=0.0188\,\text{Å}^{-3}$, this is $r_{\mathrm{WS}}=2.33\,\text{Å}$.
These radii, compared with the relevant potentials of the liquid problem,
are shown in figure~\ref{fig:potentials}. We have also explored
an alternative choice for the Wigner-Seitz radius \citep{Evans2010},
called the ``local Wigner-Seitz radius'' $r_{\mathrm{LWS}}=(4\pi Ng_{\mathrm{max}}/3)^{-1/3}$,
where $g_{\mathrm{max}}$ is the maximum of the pair correlator $g(r)$.
This quantity attempts to account for the increased density that the
positron would feel in the majority of collisions. In our case, this
results in a value of $r_{\mathrm{LWS}}=6.24\,\text{Å}$ in the dense
gas phase, and $r_{\mathrm{LWS}}=2.08\,\text{Å}$ in the liquid phase.

We have explored the choice of $r_{\mathrm{WS}}$ before in our investigation
of electrons in liquid argon, but found it to worsen the agreement
between our calculations and experimental measurements. However, at
that time we did not also apply an energy shift, which we discuss
in the following section.

\subsection{$\Delta V$}

Even as the positron velocity approaches zero, it will feel a background
energy in the presence of a liquid or dense gas \citep{Iakubov1982}.
This quantity is known as $V_{0}$ and has been obtained through a
combination of measurement and calculation for electron scattering
in various liquids, see \citep{Evans2010,Borghesani2006} and references
therein. As it is not possible to do these same experiments with positrons,
we instead investigate two different substitute values for $V_{0}$.
The first is $U_{2}(r\rightarrow0)$, which corresponds to the potential
calculated from the average of the surrounding atoms at the origin,
and the second surrogate is $V_{\mathrm{WS}}$, a Wigner-Seitz calculation
in the style of \citep{Evans2010}, which we will describe in more
detail in an upcoming paper. In short, the $V_{\mathrm{WS}}$ value
is found as the minimum energy solution for a wavefunction that satisfies
a ``spherical Bloch wave'' boundary condition. We have applied a
similar surrogate value for the potential shift when performing calculations
of electrons in liquid krypton \citep{White2018}. Note that the value
of $V_{\mathrm{WS}}$ itself depends on the value of $r_{m}$ and
we will refer to $V_{\mathrm{WS}}$ and $V_{\mathrm{LWS}}$ as the
potential shift from using the regular ($r_{\mathrm{WS}}$) and local
($r_{\mathrm{LWS}}$) Wigner-Seitz radii respectively.

\subsection{Pair correlators in helium\label{subsec:Liquid-helium-pair-correlator}}

\begin{figure}
\begin{centering}
\includegraphics[width=0.9\columnwidth]{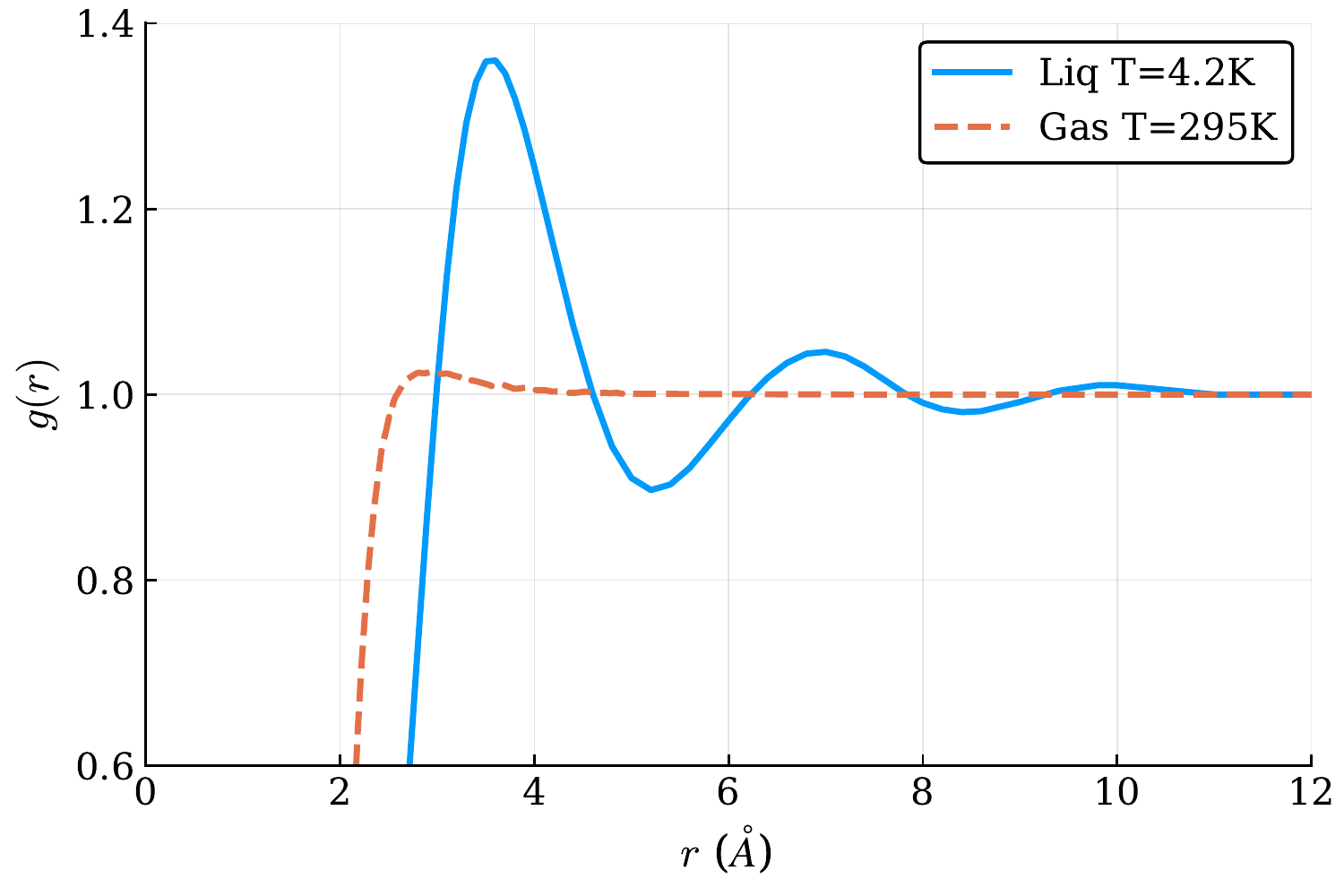}
\par\end{centering}
\caption{\label{fig:pair-corr}The pair correlators used for the liquid (solid
blue line \citep{Sears1979}) and dense gas (dashed orange line, calculated
from Monte Carlo simulations) phases.}
\end{figure}

The essential input to perform the dense fluid calculations is the
fluid pair-correlator and its Fourier transform, the static structure
factor. For liquid-phase helium at $T=4.2\ \text{K}$, we use the
pair-correlator and structure factor derived from experiments by \citep{Sears1979}.
For the dense gas case at $T=295.65$~K, we have calculated the pair-correlator
from Monte Carlo simulations with $N=10000$ atoms using an untruncated
Lennard-Jones potential with parameters \citep{Oh2013} $\epsilon_{LJ}/k_{B}=5.465$~K
and $\sigma_{LJ}=2.628\,\text{Å}$. These pair correlators are shown
in figure~\ref{fig:pair-corr}. As the pair correlator for the dense
gas is relatively flat, it can be expected that some of the dense
fluid effects will be negligible, however that contributions from
the surrounding average will still be significant.

\section{Results}

\subsection{Experimental measurements}

\begin{figure}
\begin{centering}
\includegraphics[width=0.8\columnwidth]{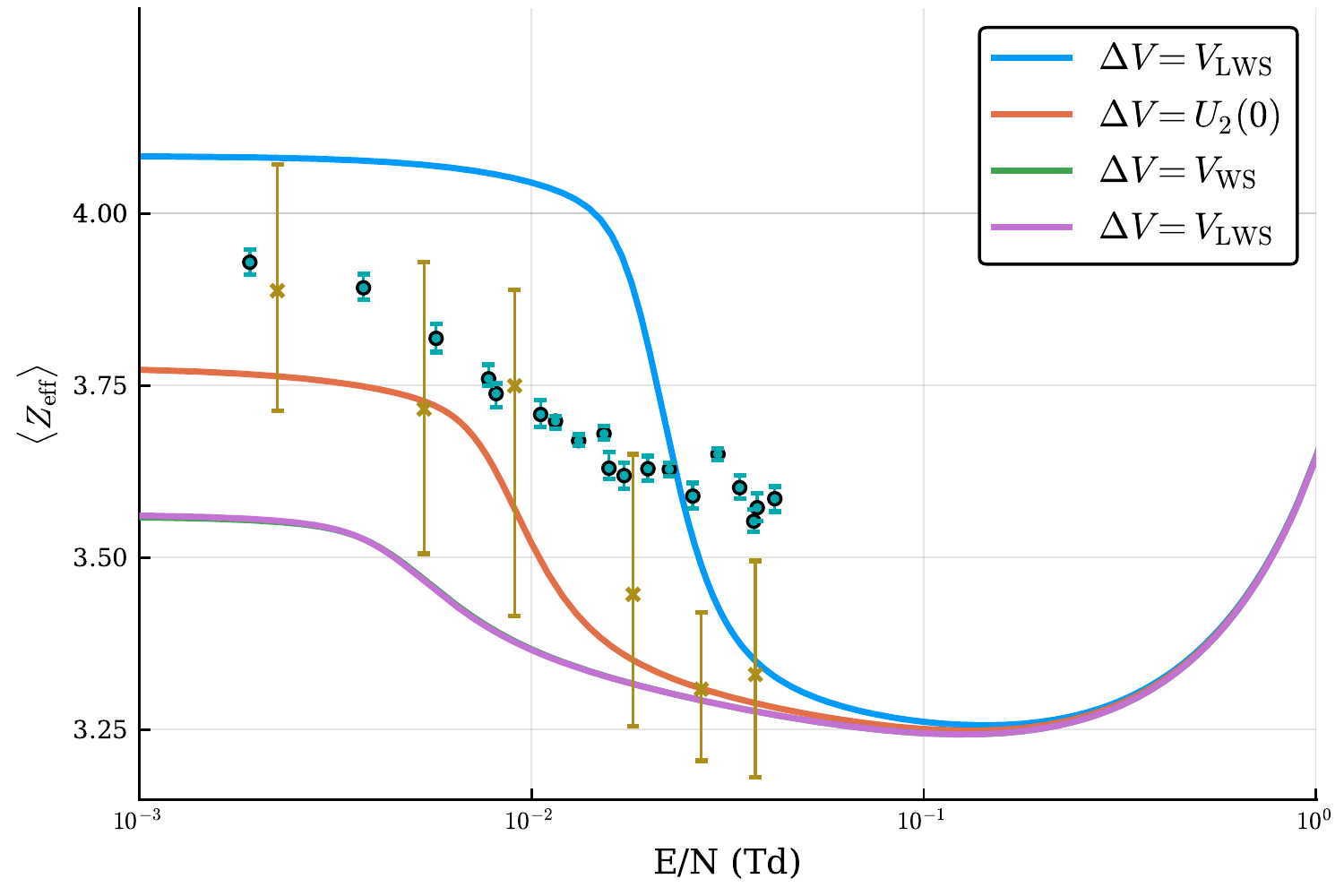}
\par\end{centering}
\caption{\label{fig:dense_gas_physical}The comparison between dense gas measurements
\citep{Davies1989} and our calculations for various physically motivated
choices of $\Delta V$. The 35.7~amg measurements are shown as filled
circles and the 3.5~amg measurements are shown for reference as crosses.
For the dense gas phase, $U_{2}(0)=-0.0084$~eV, $V_{\mathrm{WS}}=-0.0151$~eV
and $V_{\mathrm{LWS}}=-0.0150$~eV. The similarity between the $V_{\mathrm{WS}}$
and $V_{\mathrm{LWS}}$ cases is due to the negligible peak in the
pair correlator.}
\end{figure}
\begin{figure}
\begin{centering}
\includegraphics[width=0.8\columnwidth]{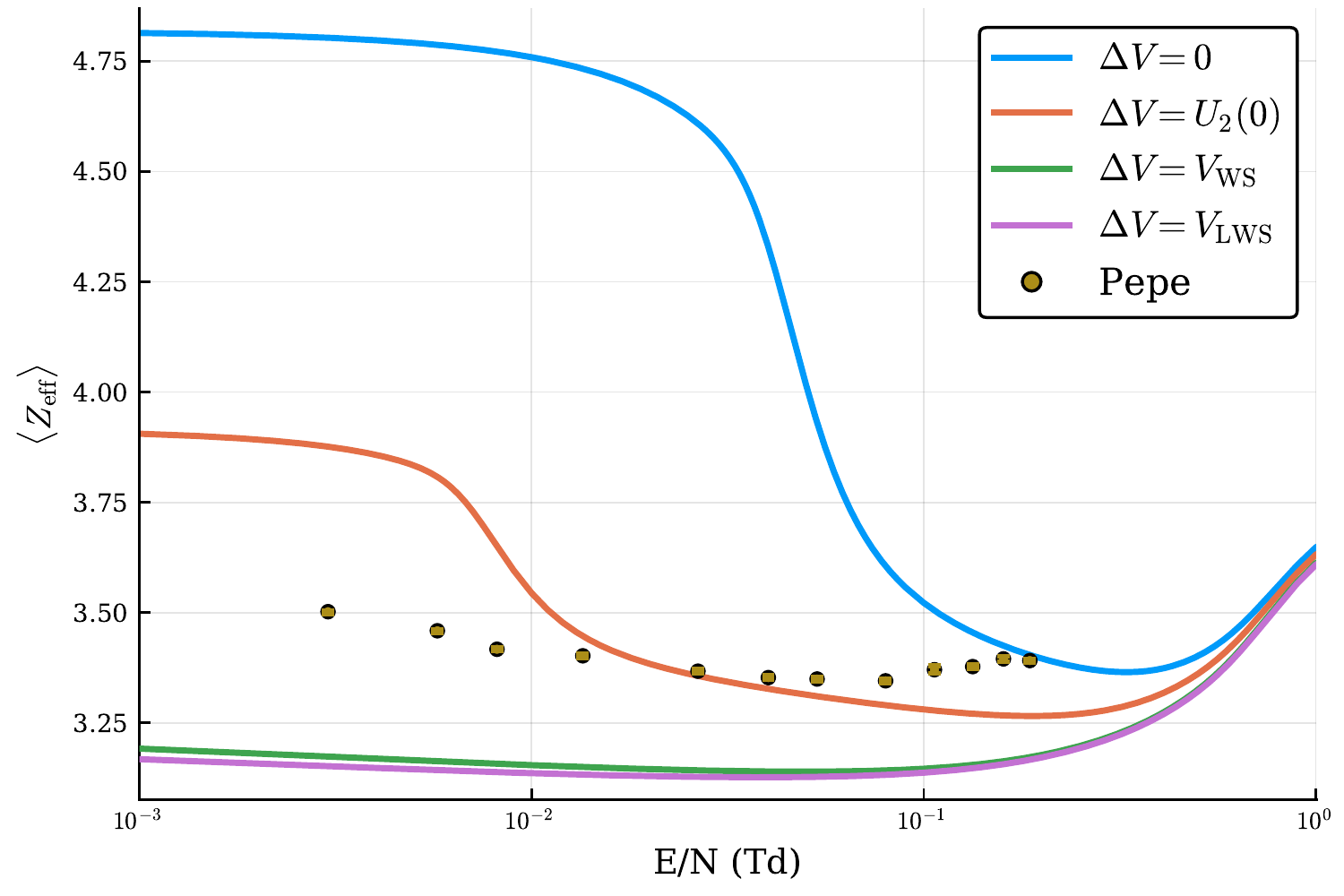}
\par\end{centering}
\caption{\label{fig:liq_no_imp}The comparison between liquid measurements
\citep{Pepe1995} and our calculations for various physically motivated
choices of $\Delta V$. For the liquid phase, $U_{2}(0)=-0.14$~eV,
$V_{\mathrm{WS}}=-0.275$~eV and $V_{\mathrm{LWS}}=-0.267$~eV. }
\end{figure}

There are several measurements of the zero-field annihilation rate,
see \citep{Fox1977,Nieminen1980,Charlton2001} for a compilation,
which allow us to assume a value of approximately $\langle Z_{\mathrm{eff}}\rangle_{T_{0}}\approx3.9$
for the dense gas phase and $\langle Z_{\mathrm{eff}}\rangle_{T_{0}}\approx3.6$
for the liquid phase. Our calculations, using several different choices
for $\Delta V$, span a range of different increases/decreases in
the zero-field $\langle Z_{\mathrm{eff}}\rangle_{T_{0}}$. In both
phases, $\Delta V=0$ shows an unusual increase in $\langle Z_{\mathrm{eff}}\rangle_{T_{0}}$
which cannot be reconciled with the experimental measurements.

We are only aware of a few measurements of the non-equilibrium field-dependent
annihilation rate. These are \citep{Davies1989} for the dense gas
phase and \citep{Pepe1995,Pepe1994} for the liquid phase. In both
cases, there is a decrease in $\langle Z_{\mathrm{eff}}\rangle(E)$
as the field is increased. While our calculations, shown in figures~\ref{fig:dense_gas_physical}
and \ref{fig:liq_no_imp}, also show a decrease it happens a) over
a larger variation of $\langle Z_{\mathrm{eff}}\rangle(E)$ for both
dense gases and liquids and b) with a shoulder at either too small
or too large a field. In addition, the calculated variation of $\langle Z_{\mathrm{eff}}\rangle(E)$
is much larger than experimentally observed. The similarity between
the $\Delta V=V_{\mathrm{WS}}$ and $\Delta V=V_{\mathrm{LWS}}$ results
in the dense gas was expected, as the key input distinguishing these
approaches is the maximum in the pair correlator, which is negligible
for the dense gas case. However, their behaviour in the liquid case
is surprising: despite a 15\% difference in $r_{m}$, the two cases
share almost identical elastic and annihilation cross sections, leading
to almost identical $\langle Z_{\mathrm{eff}}\rangle$ values.

It is also possible for us to choose different values for our simulation
parameters of $r_{m}$ and $\Delta V$, which are not necessarily
physically motivated. We have done this by scanning a wide range of
values but no particular choice allows us to obtain both the required
magnitude and field-dependence of $\langle Z_{\mathrm{eff}}\rangle(E)$,
even approximately.

While the differences between our results and the experimental measurements
in figures~\ref{fig:dense_gas_physical} and \ref{fig:liq_no_imp}
appear to be quite large, this is due to a relatively small variation
in $Z_{\mathrm{eff}}$. The differences between the $\Delta V=U_{2}(0)$
calculations and the measurements are within 5\% for the dense gas
case and 10\% for the liquid case. This could be accounted for by
assuming a systematic uncertainty in the measurements, but we will
instead consider what modifications can be made to our model to reconcile
experiment and theory in the following sections. It is worth pointing
out that the analysis involved in these measurements may be complicated
by the ordering of the lifetimes for free positron annihilation and
o-Ps annihilation: in the low density $3.5$~amagat case, o-Ps annihilation
is faster, and in the high-density $35.7$~amagat case, free positron
annihilation is faster~\citep{Charlton2020}.

We first discuss the dense gas case below in greater detail, and propose
some modifications that we can make to explain the differences. We
will then apply those considerations to the liquid phase.

\subsection{Dense gas comparison}

From figure~\ref{fig:dense_gas_physical}, we can see that the various
choices of $\Delta V$ allow us to tune the value of $\langle Z_{\mathrm{eff}}\rangle(E)$
at low fields. However, these choices all result in the same behaviour
of $\langle Z_{\mathrm{eff}}\rangle(E)$ at high fields. We believe
this should be expected from modifications due to the dense fluid,
as large kinetic energies overwhelm these effects. It is rather the
difference in the experimental measurements at higher reduced fields
between the 3.5~amagat and 35.7~amagat results that we find surprising.

We have explored some modifications to our model of the gas in order
to obtain agreement with experiment. In terms of transport quantities,
we require one or both of the following modifications: either a) an
additional source of annihilation which is significant at higher energies,
or b) a source of friction to reduce the mean energy at higher fields.
A lower mean energy has the desired side effect of increasing the
$\langle Z_{\mathrm{eff}}\rangle$ felt by the ensemble, as the annihilation
cross section is larger at lower energies.

Both of these effects can be produced by a small admixture of an impurity
in the gas. The dominant effects of a molecular species as an impurity
can be represented by two additional processes: another annihilation
pathway and an inelastic cross section. In order to separate these
effects, we first consider the zero-field case. Here, the positron
distribution (neglecting the small perturbation from annihilation)
will remain close to a thermal distribution. In this way, the additional
inelastic cross section can be neglected and only the additional annihilation
pathway will affect the measured $\langle Z_{\mathrm{eff}}\rangle_{T_{0}}$.
This leaves us with
\begin{equation}
\langle Z_{\mathrm{eff}}\rangle_{T_{0}}=\langle Z_{\mathrm{eff}}^{\mathrm{He}}\rangle_{T_{0}}+x\langle Z_{\mathrm{eff}}^{\mathrm{imp}}\rangle_{T_{0}}
\end{equation}
 where $x$ is the ratio of impurity density to helium density.

\begin{figure}
\begin{centering}
\includegraphics[width=0.9\columnwidth]{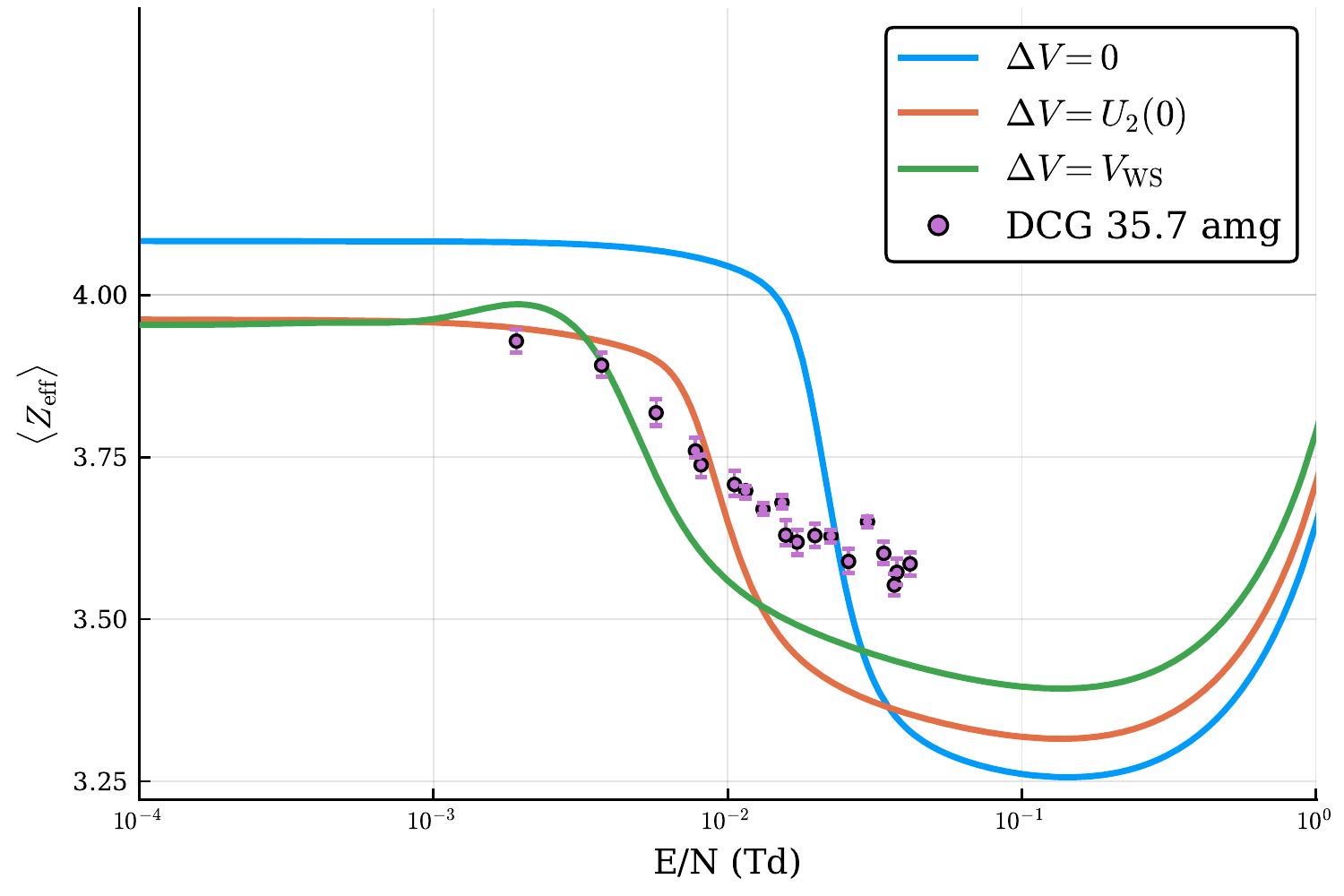}
\par\end{centering}
\caption{\label{fig:dense-imp-ann-only}The effective electron number due to
the small admixture of an impurity. The annihilation cross section
is shaped like that of ethane and the effective ethane density is
denoted as $\tilde{x}$. No additional inelastic processes are included
in these results. Note that inclusion of an impurity can only increase
the zero-field $\langle Z_{\mathrm{eff}}\rangle$, so the $\Delta V=0$
results cannot be made to match the zero-field experimental result.}
\end{figure}

As it is likely that a mix of different hydrocarbons can play the
role of impurities, we substitute their combined $\langle Z_{\mathrm{eff}}^{\mathrm{imp}}\rangle$
by a cross section that is proportional to that of ethane, i.e. $Z_{\mathrm{eff}}^{\mathrm{imp}}(\epsilon)=CZ_{\mathrm{eff}}^{\mathrm{C}_{2}\mathrm{H}_{6}}(\epsilon)$,
and reinterpret $\tilde{x}=xC$ as an \textbf{effective} ethane impurity
density. While this introduces an ambiguity into the impurity, it
removes one fitting parameter from our calculations. We emphasize
that even a few fitting parameters can allow us to fit any measured
$\langle Z_{\mathrm{eff}}\rangle(E)$, so it is important to limit
the number of these as much as possible.

Our simulations, after fitting for $\tilde{x}$ at $E=0$, are shown
in figure~\ref{fig:dense-imp-ann-only}. We can see that only a very
small admixture is required to match the experimental value of $\langle Z_{\mathrm{eff}}\rangle_{T_{0}}$.
However, in the case of $\Delta V=0$, no amount of impurity will
lower the $\langle Z_{\mathrm{eff}}\rangle_{T_{0}}$ value.

\begin{figure}
\begin{centering}
\includegraphics[width=0.9\columnwidth]{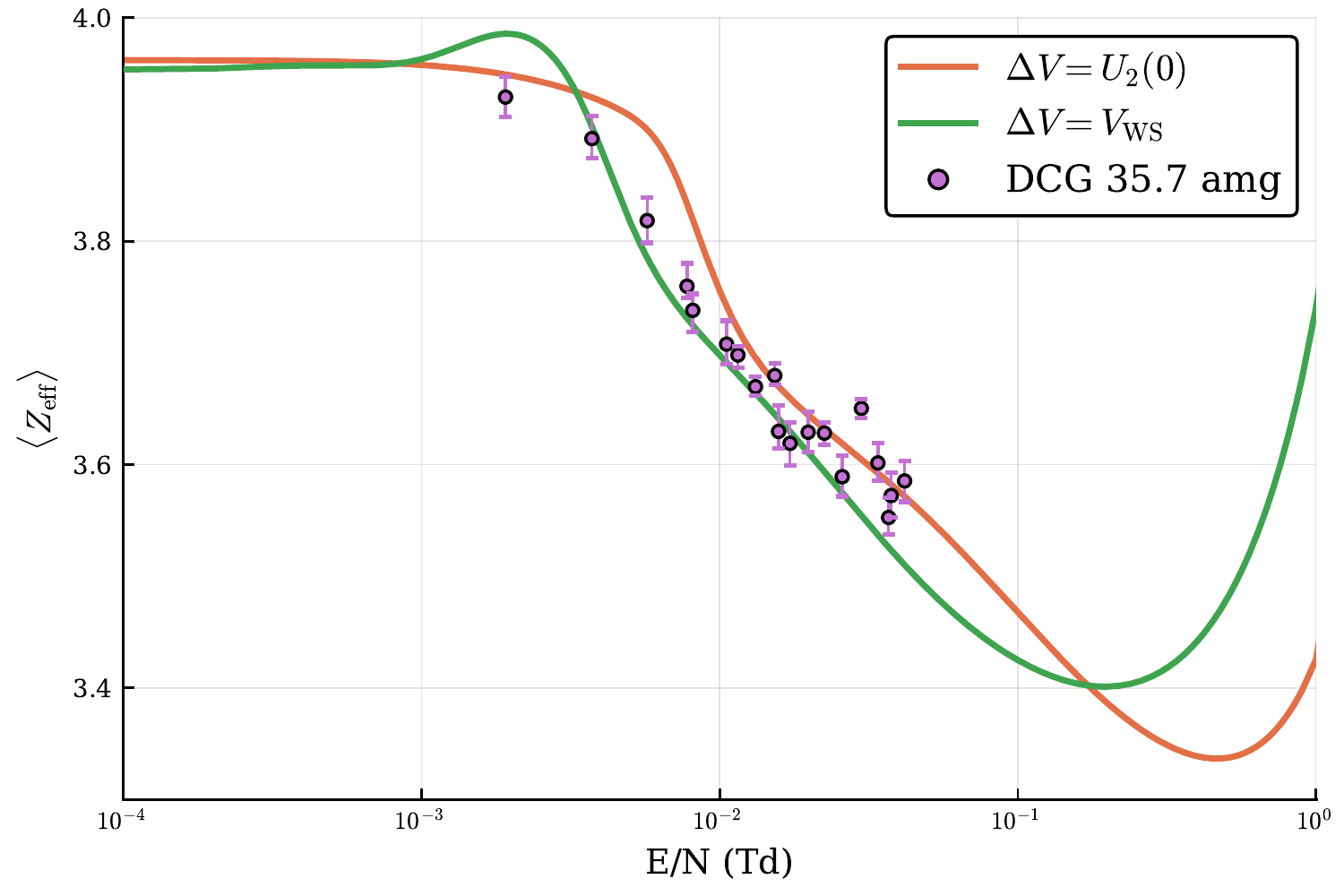}
\par\end{centering}
\caption{\label{fig:dense-imp-full}The effective electron number due to the
small admixture of an impurity, including a constant inelastic cross
section of magnitude $\tilde{A}$ and threshold $\epsilon_{\mathrm{inel}}$.
The $\Delta V=U_{2}(0)$ case corresponds to $\tilde{x}=0.067\%$,
$\tilde{A}=10^{-3}\,\text{Å}^{2}$ and $\epsilon_{\mathrm{inel}}=2.5$~eV
and the $\Delta V=V_{\mathrm{WS}}$ case corresponds to $\tilde{x}=0.15\%$,
$\tilde{A}=10^{-4}\,\text{Å}^{2}$ and $\epsilon_{\mathrm{inel}}=2$~eV}
\end{figure}

We now turn to including the second-most significant aspect of an
impurity, which is the introduction of inelastic collisions with lower
threshold energies than helium. As we again want to consider a range
of hydrocarbon impurities, we use a surrogate cross section of constant
magnitude $A$ and threshold $\epsilon_{\mathrm{inel}}$. We again
reinterpret this quantity as the magnitude $\tilde{A}=xA$ which is
an \textbf{effective} inelastic cross section, indicating a magnitude
relative to the density of helium. By doing this, there are only three
parameters to characterise the impurity: $\tilde{x}$, $\tilde{A}$
and $\epsilon_{\mathrm{inel}}$. The value of $\tilde{x}$ is fixed
by the zero-field annihilation rate, so we now vary the latter two
parameters to obtain the best fits shown in figure~\ref{fig:dense-imp-full}.

In all cases, the fits perform reasonably well and provide good agreement
over most of the range of experimental measurements. The fit for $\Delta V=V_{\mathrm{WS}}$
includes an additional additional peak at around $E/N=2\times10^{-3}$~Td
not seen in the experimental data, while the $\Delta V=U_{2}(0)$
curve does not follow the data as closely.

\begin{figure}
\begin{centering}
\includegraphics[width=0.9\columnwidth]{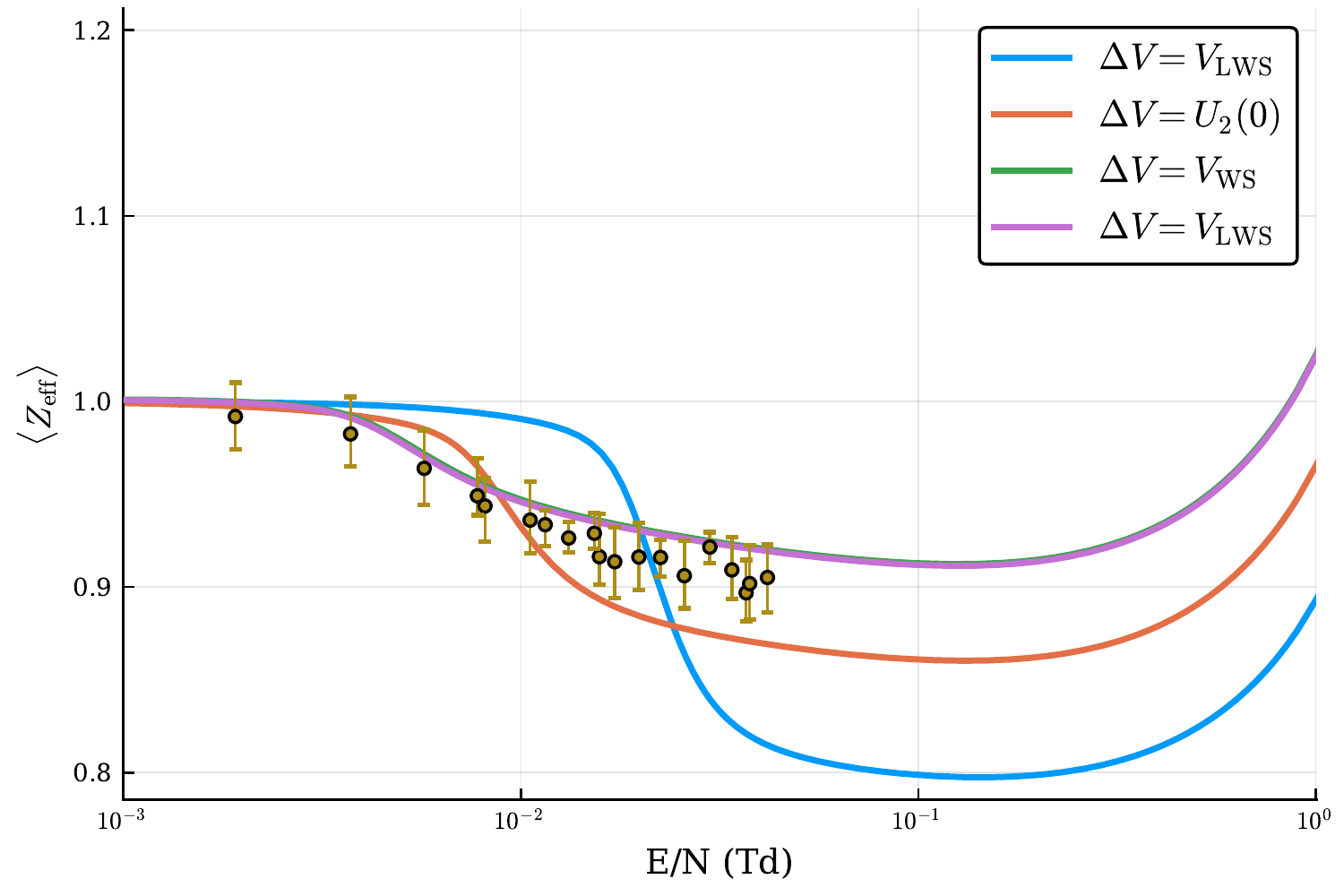}
\par\end{centering}
\caption{\label{fig:dense-phys-scaled}The effective electron number as shown
in figure~\ref{fig:dense_gas_physical} but scaled to the zero-field
value $\langle Z_{\mathrm{eff}}\rangle_{T_{0}}$. No impurity is included
in these calculations. This is useful in the scenario that the comparison
between absolute values is not possible. In this scenario, the $\Delta V=V_{\mathrm{WS}}$
and $\Delta V=V_{\mathrm{LWS}}$ cases represent the best fit.}
\end{figure}

As an alternative, we can choose to believe that the comparison of
absolute values from our calculation and measurement may not be well
posed, and instead we can compare the $\langle Z_{\mathrm{eff}}\rangle(E)$
values relative to the zero-field $\langle Z_{\mathrm{eff}}\rangle_{T_{0}}$.
This is shown in figure~\ref{fig:dense-phys-scaled}, where the $\Delta V=V_{\mathrm{WS}}$
and $\Delta V=V_{\mathrm{LWS}}$ appear to give the closest fit, although
all choices are not unreasonable.

\subsection{Liquid comparison}

As with the dense gas case, we can apply the same steps to include
an impurity to better match the experimental measurements. There is
less likelihood for the presence of an impurity in liquid helium,
as it would be expected to freeze out of the liquid. In any case,
we can consider the effect it would have.

For the liquid the $\Delta V=U_{2}(0)$ case produces a zero-field
value which is higher than the experiment, even without the inclusion
of an impurity. This means that we can only consider the $\Delta V=V_{\mathrm{WS}}$
case as suitable to add an impurity.

\begin{figure}
\begin{centering}
\includegraphics[width=0.9\columnwidth]{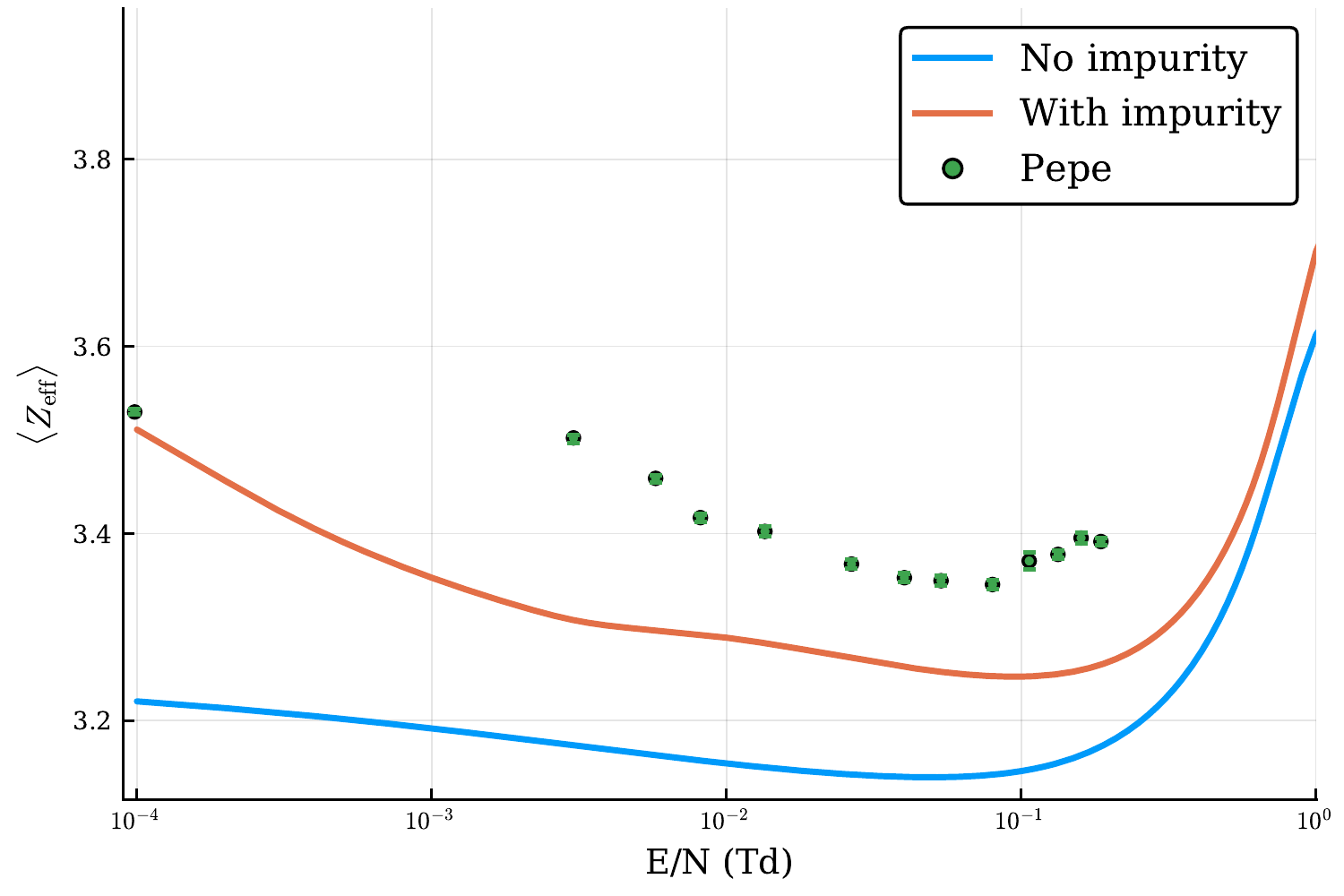}
\par\end{centering}
\caption{\label{fig:full-imp-full}The effective electron number in liquid
helium, due to the small admixture of an impurity and using $\Delta V=V_{\mathrm{WS}}$.
The impurity parameters are $\tilde{x}=0.1\%$, $\tilde{A}=10^{-4}\text{Å}^{2}$
and $\epsilon_{\mathrm{inel}}=2$~eV. Although the inclusion of the
impurity can adjust the zero-field rate to bring it into agreement
with the experimental measurement, the rest of the field range is
not in agreement, even with the inclusion of an inelastic process.
The uptick in the experimental measurements at the higher fields has
been shown to be due to Ps formation as the positrons reached a steady-state
distribution.}
\end{figure}

The effect of the impurity in the $\Delta V=V_{\mathrm{WS}}$ case
is shown in figure~\ref{fig:full-imp-full}, with and without the
additional of an inelastic process. It is clear to see that we cannot
obtain agreement. This is somewhat surprising, as we have some free
parameters to manipulate. We believe this suggests that there is a
contribution missing from our calculations, which is due to multiple
scattering at high densities.

We should also point out that we should not aim to fit the uptick
in the experimental results at high fields. This has been shown \citep{Pepe1995}
to be an apparent increase only, and is actually due to the formation
of positronium with ionised electrons. This spur-enhanced Ps formation
is estimated to be at most 1\% and only occurs at higher fields. As
the apparent $\langle Z_{\mathrm{eff}}\rangle$ is about 1.2\% larger
at the higher fields, this fits almost perfectly with this explanation.

We can also consider the possibility of positrons forming self-trapped
clusters of higher density in the helium liquid \citep{Manninen1978}.
However, these clusters have been found to only be present for densities
less than that of liquid helium. Hence, we can ignore this mechanism
as a source of increased $Z_{\mathrm{eff}}$.

Finally, we note that we have not accounted for a difference between
the applied and effective electric fields due to the permittivity
of the liquid. This is because the effect is negligible, as the dielectric
constant \citep{Chase1961} of helium is $1.05\approx1$.

\begin{figure}
\begin{centering}
\includegraphics[width=0.9\columnwidth]{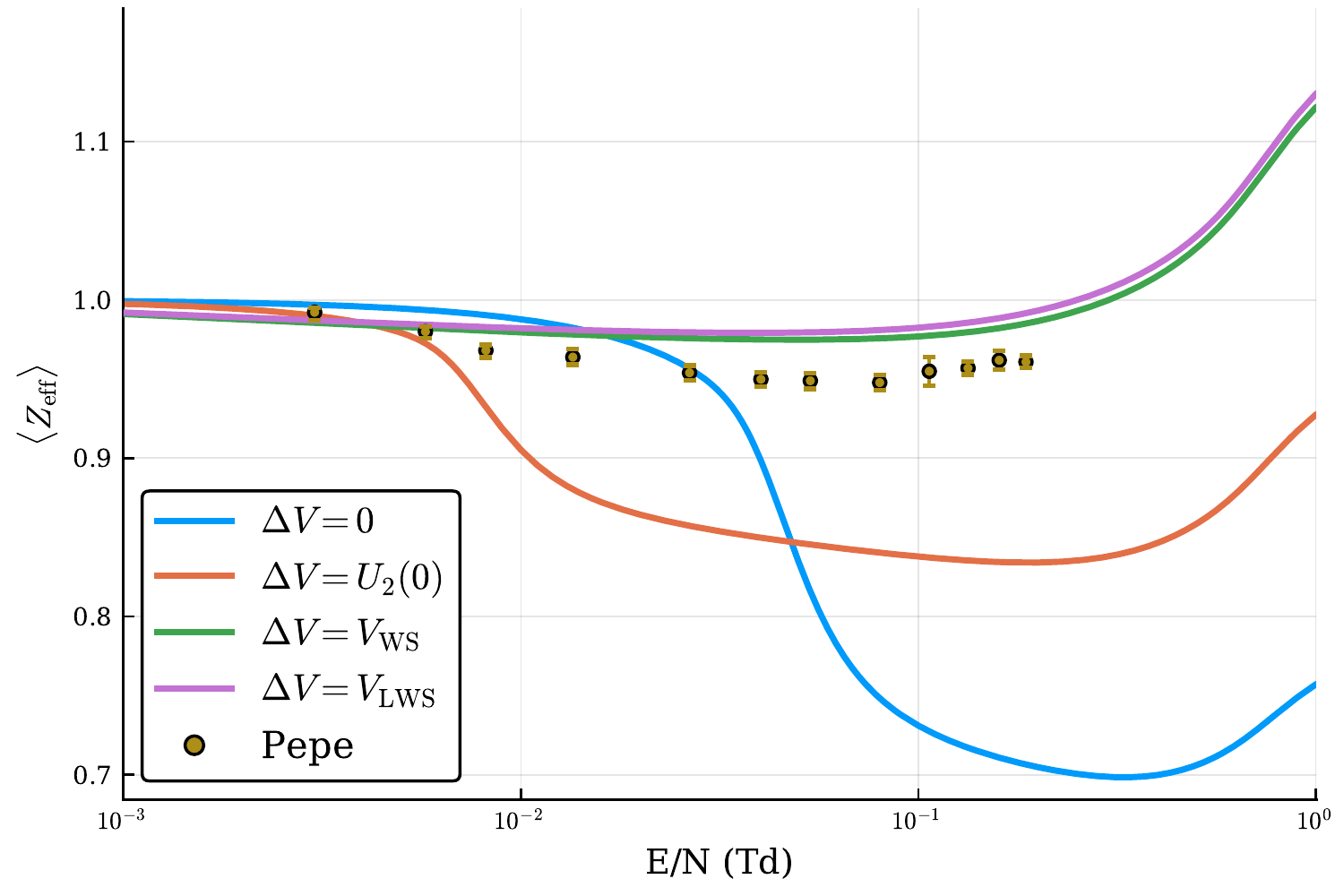}
\par\end{centering}
\caption{\label{fig:liquid-phys-scaled}The effective electron number as shown
in figure~\ref{fig:dense_gas_physical} but scaled to the zero-field
value $\langle Z_{\mathrm{eff}}\rangle(E=0)$. No impurity is included
in these calculations.}
\end{figure}

Again, we have made a comparison with the relative difference to the
zero-field $\langle Z_{\mathrm{eff}}\rangle$ value, shown in figure~\ref{fig:liquid-phys-scaled}.
In contrast to the similar comparison in figure~\ref{fig:dense-phys-scaled},
there is a much bigger difference in the choices of $\Delta V$ for
calculation, but the $\Delta V=V_{\mathrm{WS}}$ and $\Delta V=V_{\mathrm{LWS}}$
choices remain closest to the experimental measurements.

While the inability to fit the liquid results is problematic for our
calculation method, we still believe that our approach to obtain agreement
for the dense gas case is valid. This is because the density of 35.7~amg
in the measurements of \citep{Davies1989} is a rather dilute density,
so multiple-scattering effects should also be relatively weak. However,
we cannot completely rule out the possibility that our dense gas calculations
are also lacking some additional physical behaviour.

\section{Conclusions}

We have modelled the transport of positrons under an applied electric
field through dense fluids of helium and compared our predictions
of annihilation rates to experimental measurements in the dense gas
and liquid phases. Our model includes modifications due to coherent
scattering and screening of the interaction potential between the
positron and the helium atom, which have been discussed in previous
articles. This article has introduced additional considerations for
the annihilation rate due to electrons from the surrounding atoms,
and we have shown that double counting should be avoided in the averaging
process.

While our model does not provide results in complete agreement with
experimental measurements, we are able to include a very small ($\approx0.1\%$)
contribution of an impurity that is representative of a hydrocarbon
to vastly improve this agreement. The impurity is motivated by the
noticeable difference between the dilute gas and dense gas measurements
and is represented using a model which includes three fitting parameters.

Our model has been extended from our previous articles, to include
one adjustable parameter, $\Delta V$, for which we have explored
three physically-motivated values: a) $\Delta V=U_{2}(0)$, b) $\Delta V=V_{\mathrm{WS}}$
and c) $\Delta V=V_{\mathrm{LWS}}$. These values are a) the total
polarisation potential of the surrounding atoms at the origin of the
focus atom and b) the ground state energy of the conduction band,
calculated in a Wigner-Seitz model and c) a calculation in the Wigner-Seitz
model using a ``local'' Wigner-Seitz radius \citep{Evans2010}.
In both the dense gas and the liquid, the effect of $\Delta V=V_{\mathrm{WS}}$
and $\Delta V=V_{\mathrm{LWS}}$ were found to be nearly identical.
In the case of a dense gas of helium, any of these choices can be
made to agree with the experimental measurements, using different
choices of an impurity admixture. However, for the case of liquid
helium, only the choices of $\Delta V=V_{\mathrm{WS}}$ or $\Delta V=V_{\mathrm{LWS}}$
were found to be compatible, yet there remained significant discrepancies
between our calculated values and the experimental measurements.

Our results, using the Boltzmann equation description outlined in
this article, have also been independently verified using a Monte-Carlo
calculation. Details of that implementation are available in \citep{Tattersall2017,Tattersall2015}.

One of the reasons that impurities can play a large role in our current
investigations is due to the very small $Z_{\mathrm{eff}}$ of helium.
In the future, we wish to model positron transport through fluids
of larger atomic species. These atoms, with many more electrons, may
provide a means to better test our calculations by suppressing the
potential effects of impurities.

In addition, we wish to explore further choices of $\Delta V$ and
determine a method to uniquely specify its value. One manner in which
to do this is to consider species in which there are a larger range
of densities with experimental measurements, such as krypton \citep{Jacobsen1986}.
We also intend to include further multiple-scattering corrections
to both $\Delta V$ and $Z_{\mathrm{eff}}$ \citep{Ziman1966,Lax1951},
to see if these can identify the current disagreement between our
calculations and experimental measurement.

\section{Acknowledgments}

We would like to acknowledge valuable discussions with Prof. M. Charlton
regarding the possibility of impurities in the dense gas experimental
measurements.

\bibliographystyle{danny}
\bibliography{PositronsLiquidHelium,library,library_extra}

\begin{thebibliography}{10}

\bibitem{Charlton2001}
M.~Charlton and J.~W. Humberston.
\newblock \emph{{Positron Physics}}.
\newblock Cambridge University Press, Cambrigde, 2001.

\bibitem{Prantzos2011}
N.~Prantzos, C.~Boehm, A.~M. Bykov, et~al.
\newblock \emph{{The 511 keV emission from positron annihilation in the
  Galaxy}}.
\newblock Rev. Mod. Phys., \textbf{83} 1001, 2011.

\bibitem{RobsonBook}
R.~Robson, R.~White, and M.~Hildebrandt.
\newblock \emph{{Fundamentals of Charged Particle Transport in Gases and
  Condensed Matter}}.
\newblock CRC Press, London, 2017.

\bibitem{Lekner1967a}
J.~Lekner.
\newblock \emph{{Motion of electrons in liquid argon}}.
\newblock Physical Review, \textbf{158} 130, 1967.

\bibitem{Boyle2015}
G.~J. Boyle, R.~P. McEachran, D.~G. Cocks, et~al.
\newblock \emph{{Electron scattering and transport in liquid argon}}.
\newblock Journal of Chemical Physics, \textbf{142} 154507, 2015.

\bibitem{Boyle2016}
G.~J. Boyle, R.~P. McEachran, D.~G. Cocks, et~al.
\newblock \emph{{Ab initio electron scattering cross-sections and transport in
  liquid xenon}}.
\newblock Journal of Physics D: Applied Physics, \textbf{49} 355201, 2016.

\bibitem{White2018}
R.~D. White, D.~Cocks, G.~Boyle, et~al.
\newblock \emph{{Electron transport in biomolecular gaseous and liquid systems:
  theory, experiment and self-consistent cross-sections}}.
\newblock Plasma Sources Science and Technology, \textbf{27} 053001, 2018.

\bibitem{Mizogawa1985}
T.~Mizogawa, Y.~Nakayama, T.~Kawaratani, et~al.
\newblock \emph{{Precise measurements of positron-helium total cross sections
  from 0.6 to 22 eV}}.
\newblock Physical Review A, \textbf{31} 2171, 1985.

\bibitem{Sullivan2008}
J.~P. Sullivan, C.~Makochekanwa, A.~Jones, et~al.
\newblock \emph{{High-resolution, low-energy positron scattering from helium:
  Measurements of the total scattering cross section}}.
\newblock Journal of Physics B: Atomic, Molecular and Optical Physics,
  \textbf{41}, 2008.

\bibitem{Davies1989}
S.~A. Davies, M.~Charlton, and T.~C. Griffith.
\newblock \emph{{Free positron annihilation in gases under the influence of a
  static electric field}}.
\newblock J. Phys. B, \textbf{22} 327, 1989.

\bibitem{Pepe1995}
I.~Pepe, D.~A.~L. Paul, J.~Steyaert, et~al.
\newblock \emph{{Positron annihilation in liquid helium and liquid argon under
  an electric field}}, 1995.

\bibitem{Boyle2014}
G.~Boyle, M.~Casey, R.~White, et~al.
\newblock \emph{{Transport theory for low-energy positron thermalization and
  annihilation in helium}}.
\newblock Phys. Rev. A, \textbf{89} 022712, 2014.

\bibitem{McEachran2019}
R.~P. McEachran and A.~D. Stauffer.
\newblock \emph{{Positron scattering from helium}}.
\newblock Journal of Physics B: Atomic, Molecular and Optical Physics,
  \textbf{52} 115203, 2019.

\bibitem{McEachran1977}
R.~P. McEachran, D.~L. Morgan, A.~G. Ryman, et~al.
\newblock \emph{{Positron scattering from noble gases}}.
\newblock Journal of Physics B: Atomic and Molecular Physics, \textbf{10} 663,
  1977.

\bibitem{McEachran1977corr}
R.~P. McEachran, D.~L. Morgan, A.~G. Ryman, et~al.
\newblock Journal of Physics B: Atomic and Molecular Physics, \textbf{11} 951,
  1978.

\bibitem{Charlton1999}
M.~Charlton.
\newblock \emph{{Experimental studies of positrons scattering in gases}}.
\newblock Reports on Progress in Physics, \textbf{48} 737, 1999.

\bibitem{McEachran1978}
R.~P. McEachran, A.~G. Ryman, and A.~D. Stauffer.
\newblock \emph{{Positron scattering from neon}}.
\newblock Journal of Physics B: Atomic and Molecular Physics, \textbf{11} 551,
  1978.

\bibitem{Paul1957}
D.~Paul and R.~Graham.
\newblock \emph{{Annihilation of Positrons in Liquid Helium}}.
\newblock Physical Review, \textbf{106} 16, 1957.

\bibitem{Charlton2009}
M.~Charlton.
\newblock \emph{{Positron transport in gases}}.
\newblock Journal of Physics: Conference Series, \textbf{162} 012003, 2009.

\bibitem{Green2017}
D.~G. Green.
\newblock \emph{{Positron Cooling and Annihilation in Noble Gases}}.
\newblock Physical Review Letters, \textbf{119} 2, 2017.

\bibitem{Cohen1967}
M.~H. Cohen and J.~Lekner.
\newblock \emph{{Theory of hot electrons in gases, liquids and solids}}.
\newblock Physical Review, \textbf{158} 305, 1967.

\bibitem{Evans2010}
C.~M. Evans and G.~L. Findley.
\newblock \emph{{Energy of the conduction band in near critical point fluids}}.
\newblock Physics Research International, \textbf{2010}, 2010.

\bibitem{Iakubov1982}
I.~T. Iakubov and a.~G. Khrapak.
\newblock \emph{{Self-trapped states of positrons and positronium in dense
  gases in liquids}}.
\newblock Reports on Progress in Physics, \textbf{45} 697, 1982.

\bibitem{Borghesani2006}
A.~F. Borghesani.
\newblock \emph{{Electron and ion transport in dense rare gases}}.
\newblock IEEE Transactions on Dielectrics and Electrical Insulation,
  \textbf{13} 492, 2006.

\bibitem{Sears1979}
V.~Sears, E.~Svensson, A.~Woods, et~al.
\newblock \emph{{The static structure factor and pair correlation function for
  liquid 4 He at saturated vapour pressure}}.
\newblock Technical Report November, Atomic Energy of Canada Ltd, 1979.

\bibitem{Oh2013}
S.-K. Oh.
\newblock \emph{{Modified Lennard-Jones Potentials with a Reduced
  Temperature-Correction Parameter for Calculating Thermodynamic and Transport
  Properties: Noble Gases and Their Mixtures (He, Ne, Ar, Kr, and Xe)}}.
\newblock Journal of Thermodynamics, \textbf{2013} 1, 2013.

\bibitem{Fox1977}
R.~Fox, K.~Canter, and M.~Fishbein.
\newblock \emph{{Positron and orthopositronium decay rates in helium at high
  densities}}.
\newblock Physical Review A, \textbf{15} 1340, 1977.

\bibitem{Nieminen1980}
R.~M. Nieminen.
\newblock \emph{{Nonlinear density dependence of the positron decay rate in
  helium}}.
\newblock Physical Review A, \textbf{21} 1347, 1980.

\bibitem{Pepe1994}
I.~Pepe, D.~A.~L. Paul, F.~Gimeno-Nogues, et~al.
\newblock \emph{{Positron annihilation in helium, argon and nitrogen liquids
  under an electric field}}.
\newblock Hyperfine Interactions, \textbf{89} 425, 1994.

\bibitem{Charlton2020}
M.~Charlton.
\newblock Private communication, 2020.

\bibitem{Manninen1978}
M.~Manninen and P.~Hautoj{\"{a}}rvi.
\newblock \emph{{Clustering of atoms around the positron and positive ions in
  gaseous He, Ne, and Ar}}.
\newblock Physical Review B, \textbf{17} 2129, 1978.

\bibitem{Chase1961}
C.~E. Chase, E.~Maxwell, and W.~E. Millett.
\newblock \emph{{The dielectric constant of liquid helium}}.
\newblock Physica, \textbf{27} 1129, 1961.

\bibitem{Tattersall2017}
W.~J. Tattersall, D.~G. Cocks, G.~J. Boyle, et~al.
\newblock \emph{Spatial profiles of positrons injected at low energies into
  water: influence of cross section models}.
\newblock Plasma Sources Science and Technology, \textbf{26} 045010, 2017.

\bibitem{Tattersall2015}
W.~J. Tattersall, D.~G. Cocks, G.~J. Boyle, et~al.
\newblock \emph{Monte carlo study of coherent scattering effects of low-energy
  charged particle transport in percus-yevick liquids}.
\newblock Phys. Rev. E, \textbf{91} 043304, 2015.

\bibitem{Jacobsen1986}
F.~M. Jacobsen, N.~Gee, and G.~R. Freeman.
\newblock \emph{{Electron mobility in liquid krypton as function of density,
  temperature, and electric field strength}}.
\newblock Physical Review A, \textbf{34} 2329, 1986.

\bibitem{Ziman1966}
J.~M. Ziman.
\newblock \emph{{Wave propagation through an assembly of spheres I. the
  Greenian method of the theory of metals}}.
\newblock Proceedings of the Physical Society, \textbf{88} 387, 1966.

\bibitem{Lax1951}
M.~Lax.
\newblock \emph{{Multiple Scattering of Waves}}.
\newblock Reviews of Modern Physics, \textbf{23} 287, 1951.

\end{thebibliography}

\end{document}